\documentclass[pra,reprint,twocolumn,showpacs]{revtex4-1}
\usepackage{epsfig,graphicx}
\usepackage{amsmath}
\usepackage{color}
\usepackage{subfigure}

\newcommand{\bra}[1]{\langle\, {#1} \,|}
\newcommand{\ket}[1]{|\, {#1} \,\rangle}
\newcommand{\braket}[2]{ \langle\, {#1} \, |\, {#2} \,\rangle }

\begin{document}

\title{Dipolar--Induced Resonance for Ultracold Bosons in a Quasi--1D Optical Lattice}
\author{N. Bartolo${}^{1,2}$, D.J. Papoular${}^1$, L. Barbiero${}^{3,4}$, C. Menotti${}^1$, A. Recati${}^1$} 
\affiliation{${}^{1}$INO-CNR BEC Center and Dipartimento di Fisica, Universit\`a di Trento, 38123 Povo, Italy\\
${}^{2}$Laboratoire Charles Coulomb, CNRS, Universit\'e Montpellier 2, France\\
${}^{3}$Dipartimento di Fisica e Astronomia ``Galileo Galilei'', Universita di Padova, Italy\\
${}^{4}$Laboratoire de Physique Th\'eorique, CNRS, Universit\'e de Toulouse, France
}

\begin{abstract}
We study the role of the Dipolar--Induced Resonance (DIR) in a quasi--one--dimensional system of ultracold bosons.
We first describe the effect of the DIR on two particles in a harmonic trap. Then, we consider a deep optical lattice loaded with ultracold dipolar bosons. In order to describe this system, we introduce a novel atom--dimer extended Bose--Hubbard model, which is the minimal model correctly accounting for the DIR. We analyze the impact of the DIR on the phase diagram at $T\!=\!0$ by exact diagonalization of a small--sized system. We show that the DIR strongly affects this phase diagram. In particular, we predict the mass density wave to occur in a narrow domain corresponding to weak nearest--neighbor interactions, and the occurrence of a collapse phase for stronger dipolar interactions. 
\end{abstract}


\maketitle

\section*{Introduction}
The recent experimental developments in the field of ultracold dipolar gases have opened up fascinating prospects for the study of systems exhibiting Dipole--Dipole Interaction (DDI) \cite{lahaye_RPP2009,baranov_ChemRev2012}. Bose--Einstein Condensates (BECs) of magnetic atoms have been realized using Chromium \cite{griesmaier_PRL2005}, 
Erbium \cite{aikawa_PRL2012}, and Dysprosium \cite{lu_PRL2011}. However, atomic magnetic moments are small ($\lesssim 10\,\mu_\mathrm{B}$, where $\mu_\mathrm{B}$ is the Bohr magneton), and therefore the effects of the DDI observed with these systems have remained perturbative up to now \cite{stuhler_PRL2005}. The recent realization of the ultracold heteronuclear molecules RbK \cite{ni_Science2008} and NaK \cite{wu_PRL2012}, which both carry electric dipole moments of the order of $1$ Debye, offer a promising route towards stronger DDI effects, but quantum degeneracy still remains to be achieved with these systems. Rydberg atoms boast much larger dipole moments but yield challenging experimental problems associated with time and length scales \cite{saffman_RMP2010}.

The DDI is anisotropic and long--ranged, and dipolar gases thus allow for the quantum simulation of more general Hamiltonians than those accessible with non--dipolar neutral particles, whose interaction is described by the standard $s$--wave interaction \cite{bloch_NatPhys2012}. Trapping a dipolar system into lower dimensions stabilizes it  with respect to two--body \cite{deMiranda_NatPhys2011} and many--body \cite{fischer_PRA2006} instabilities caused by the attractive part of the 3D DDI. This has prompted detailed studies of dipolar systems in 2D and quasi--2D \cite{matveeva_PRL2012,*lu_PRA2012,*babadi_PRB2011,*babadi_arxiv2012,sowinski_PRL2012},
bilayer \cite{pikovski_PRL2010,*zinner_PRA2012,*klawunn_arxiv2011,*chan_PRA2010}, and 
quasi--1D \cite{sinha_PRL2007,deuretzbacher_PRA2010,*deuretzbacher_PRA2013_err,maluckov_PRL2012,zinner_NJP2013} geometries.

Experiments involving dipolar bosons in optical lattices have recently been performed both with atomic BECs \cite{pasquiou_PRL2011} and non--condensed dipolar molecules \cite{chotia_PRL2012,yan_arxiv2013}. Up to now, their standard theoretical description has relied on the Extended Bose--Hubbard Model (EBHM) accounting for the interaction between nearest and more distant neighbors
\cite{lahaye_RPP2009}.
The 1D EBHM has revealed the occurrence, beyond the standard Mott--Insulator (MI) and superfluid (SF) phases, of a Mass Density Wave (MDW) phase \cite{pai_PRB2005,maluckov_PRL2012} and a Haldane Insulator phase \cite{dallaTorre_PRL2006,rossini_NJP2012} .

The proper description of specific atomic systems by lattice models such as the EBHM requires a careful mapping between models and physical systems.
This has already been
pointed out for  the Hubbard
model \cite{buchler_PRL2010}, 
but the  non--trivial  effects associated  with long--range  and
anisotropic   interactions   are   even  more   important.
As a first step in this direction, we analyze the important role played by the Dipolar--Induced Resonance (DIR)  \cite{marinescu_PRL1998,shi_PRA2012}, which is a low--energy resonance occurring when the dipole strength is varied. We show that the DIR affects both the two--body and the many--body physics of the system (see e.g.~\cite{qi_PRL2013} about the BEC---BCS crossover).

In this article, we consider a quasi--1D lattice system of bosonic dipoles in the tight--binding regime \footnote{Our analysis holds for both atomic and molecular dipoles.}  \cite{trefzger_JPB2011,bloch_RMP2008}. We assume that the dipole moments are aligned perpendicularly to the trap axis. In this situation, a single DIR occurs. Accounting for it requires going beyond the single--band EBHM. We develop a novel atom--dimer extended Bose--Hubbard model, which is the minimal model capturing the DIR. Even at this level, we find that the DIR has a strong impact on the many--body phase diagram as compared to previous descriptions \cite{pai_PRB2005,rossini_NJP2012}.

The scattering and bound--state properties of the DDI have been studied numerically for free--space models \cite{kanjilal_PRA2008}, and for 3D and 2D lattice systems \cite{hanna_PRA2012}. In our quasi--1D geometry, we model the DDI using an effective potential obtained by averaging the transverse degrees of freedom into the harmonic--oscillator ground state \cite{sinha_PRL2007,deuretzbacher_PRA2010,*deuretzbacher_PRA2013_err}:
\begin{equation} \label{eq:V1D}
\begin{aligned}
V&_\mathrm{1D}(x)=
\left(g_{1D}-\frac{\hbar^2}{m}\frac{2\,r^*}{3\,l_\perp^2}\right)\delta(x)\\
&+\frac{\hbar^2}{m}\frac{r^*}{l_\perp^3} \left[
\sqrt{\frac{\pi}{8}}\,e^{\frac{1}{2}\frac{x^2}{l_\perp^2}}
\left(\frac{x^2}{l_\perp^2}+1\right) \text{Erfc}\left(\frac{|x|}{\sqrt{2}l_\perp}\right)
-\frac{|x|}{2\,l_\perp}
\right]
\ ,
\end{aligned}
\end{equation}
where  
$r^*=mD^2/\hbar^2$ is the dipolar length, with $D$ being the dipolar strength.
The range of this potential is determined by the oscillator length $l_\perp=(\hbar/m\omega_\perp)^{1/2}$ in the strongly--confined directions 
$y$ and $z$. 
The term $g_{1D}=2\hbar^2a_\mathrm{3D}/ml_\perp^2$ is the strength of the 
$s$--wave contact interaction for a 3D scattering length $a_\mathrm{3D}$ \cite{petrov_PRL2000}, 
which can be manipulated using a Feshbach resonance \cite{chin_RMP2010}.
It competes with the DDI to determine the stability and the 
phase of the system \cite{koch_NatPhys2008}. We assume 
$g_\mathrm{1D}=0$ unless otherwise specified. Under this assumption, Eq.~(\ref{eq:V1D}) still contains a contact term proportional to $r^*$. The use of Eq.~(\ref{eq:V1D}) amounts to neglecting the role of confinement--induced resonances \cite{olshanii_PRL1998}. Their interplay with the DIR might lead to even richer physics, which we are currently investigating.

\section{Two--body physics} 
The basic building block of our many--body lattice Hamiltonian (Eq.~(\ref{eq:atdimEBHM})) is provided by the solution of the two--body problem in a single lattice site. Hence, we solve for the ground--state of two dipolar bosons in a 1D harmonic well, with the trapping frequency $\omega_0$ and the oscillator length $l_0=(\hbar/m\omega_0)^{1/2}$.
The center--of--mass and relative motions are decoupled, and the relative motion is governed by the Hamiltonian:
\begin{equation}
  H_\mathrm{2B}=\frac{p^2}{2m_r}+
  \frac{1}{2}m_r\omega_0^2 x^2 + V_\mathrm{1D}(x)
  \ ,
\end{equation}
where $x$ is the interparticle distance, $p$ is its conjugate momentum, and $m_r=m/2$ is the reduced mass.

\begin{figure}
  \begin{center}
    \includegraphics[width=.31\textwidth]
    {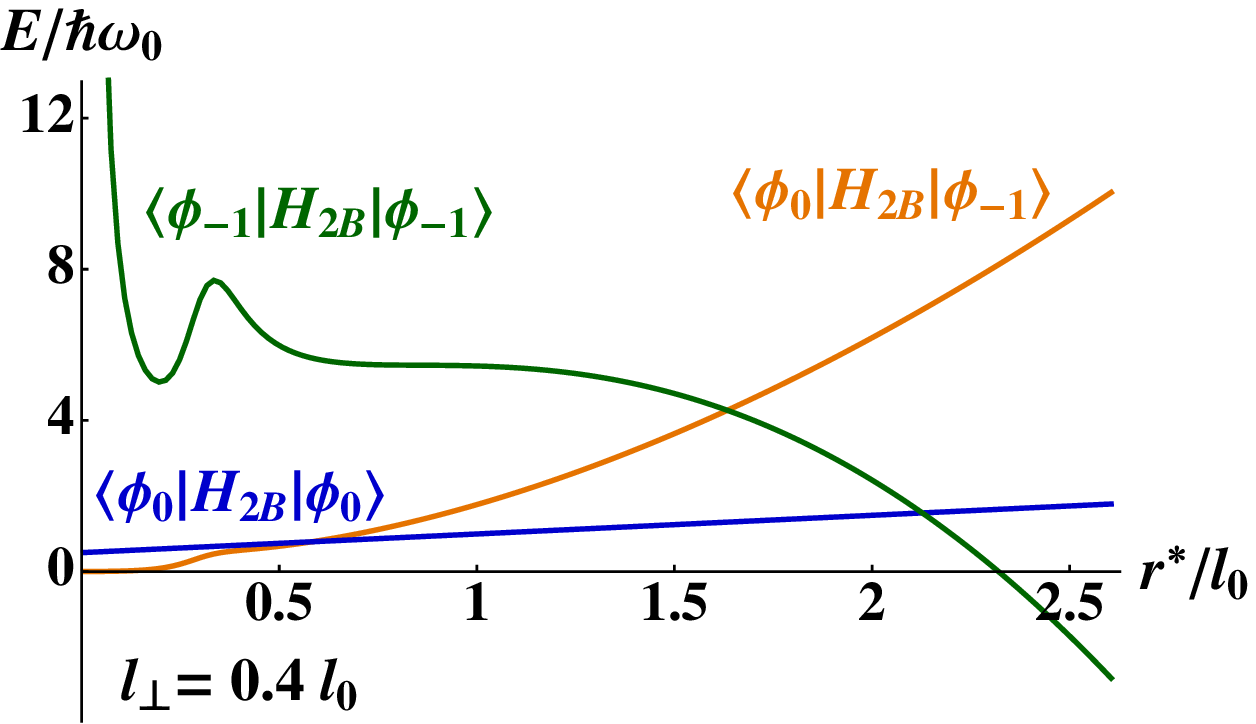}
  \end{center}
  \begin{center}
    \includegraphics[width=.31\textwidth]
    {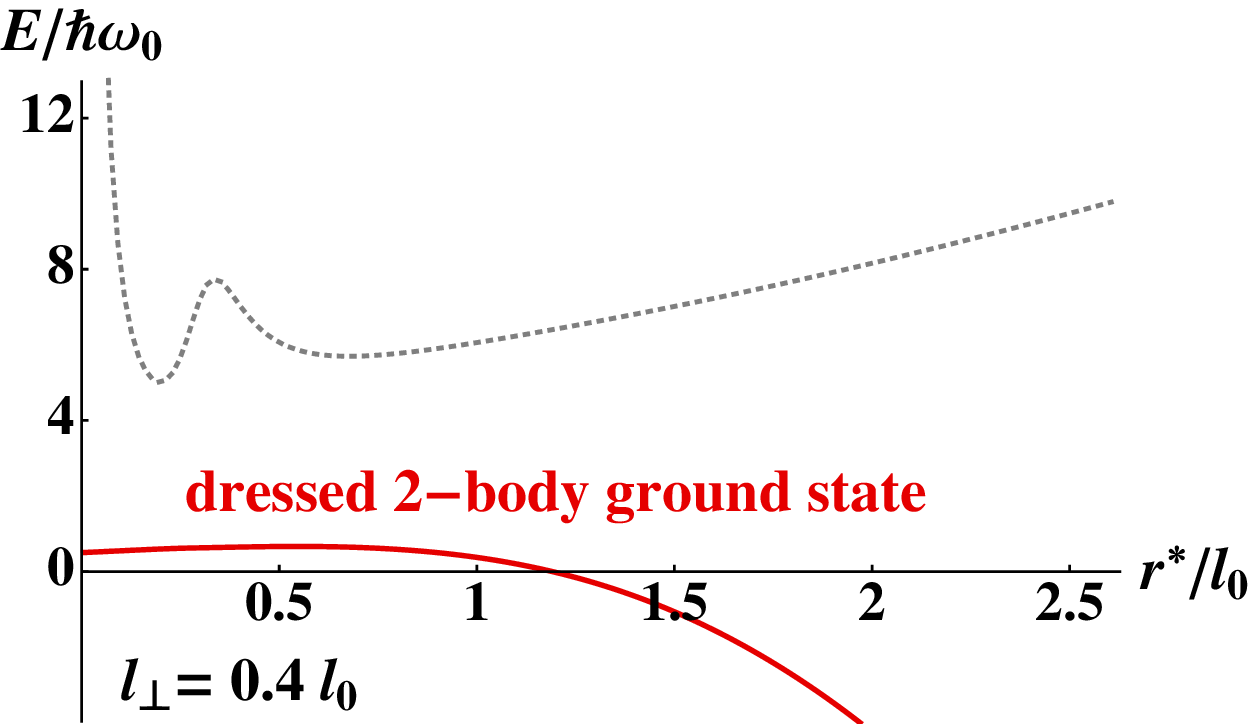}
  \end{center}
  \caption{ \label{fig:twostate} (Color online).
    Top: matrix elements of the two--state Hamiltonian $H_\mathrm{2state}$ 
    describing two bosonic dipoles in a harmonic trap, as a function of
    the dipolar length $r^*$.
    Bottom: corresponding ground--state (red) and excited--state (dashed gray)
    energies.
    }
\end{figure}
Unlike for the contact interaction \cite{busch_FoundPhys1998}, the Hamiltonian $H_\mathrm{2B}$ cannot be diagonalized analytically. We seek its ground state numerically, by considering the restriction of $H_\mathrm{2B}$ onto a subspace spanned by a finite number of basis states $\{ \ket{\phi_n} \}$. 
Depending on the value of $r^*$, $V_\mathrm{1D}$ supports either no bound state or a single one. The bound state is present for large enough values of $r^*$, and its entrance coincides with the occurrence of the DIR. The `bare' bound state supported by the attractive contact part of $V_\mathrm{1D}(x)$ plays a key role. Its wavefunction is
$\psi_\mathrm{BS}(x)=\sqrt{\kappa}\exp(-\kappa |x|)$, where $\kappa=r^*/(3l_\perp^2)$, and its cusp at $x=0$ cannot be reproduced by projecting $\ket{\psi_\mathrm{BS}}$ onto any finite number of harmonic oscillator eigenstates which are all smooth at $x=0$.
Hence, the DIR physics can only be captured if a wavefunction which has a cusp at $x=0$ is included in the basis $\{ \ket{\phi_n} \}$. 
The smallest such basis 
is $\{ \ket{\phi_0}, \ket{\phi_{-1}} \}$, where $\ket{\phi_0}$ is the ground state of the 1D harmonic oscillator with frequency $\omega_0$, and 
$\ket{\phi_{-1}}\propto 
\ket{\psi_\mathrm{BS}} - \braket{\phi_0}{\psi_\mathrm{BS}}\ket{\phi_0}$ 
is a linear combination of $\ket{\psi_\mathrm{BS}}$ and $\ket{\phi_0}$ chosen such that the basis is orthonormal. 
Hence, for a given value of $r^*$, we replace $H_\mathrm{2B}$ by the two--state Hamiltonian:
\begin{equation} \label{eq:twostate}
  H_\mathrm{2state}= 
  \begin{pmatrix}
    \bra{\phi_0} H_\mathrm{2B} \ket{\phi_0} &
    \bra{\phi_0} H_\mathrm{2B} \ket{\phi_{-1}}\\
    \bra{\phi_{-1}} H_\mathrm{2B} \ket{\phi_0} &
    \bra{\phi_{-1}} H_\mathrm{2B} \ket{\phi_{-1}} 
  \end{pmatrix}
  \ .
\end{equation}
The diagonalization of $H_\mathrm{2state}$ yields the ground--state energy $E_\mathrm{2B}(r^*)$ and wavefunction $\ket{\Psi_\mathrm{2B}(r^*)}$. 

\begin{figure}
  \begin{center}
    \includegraphics[width=.3\textwidth]
    {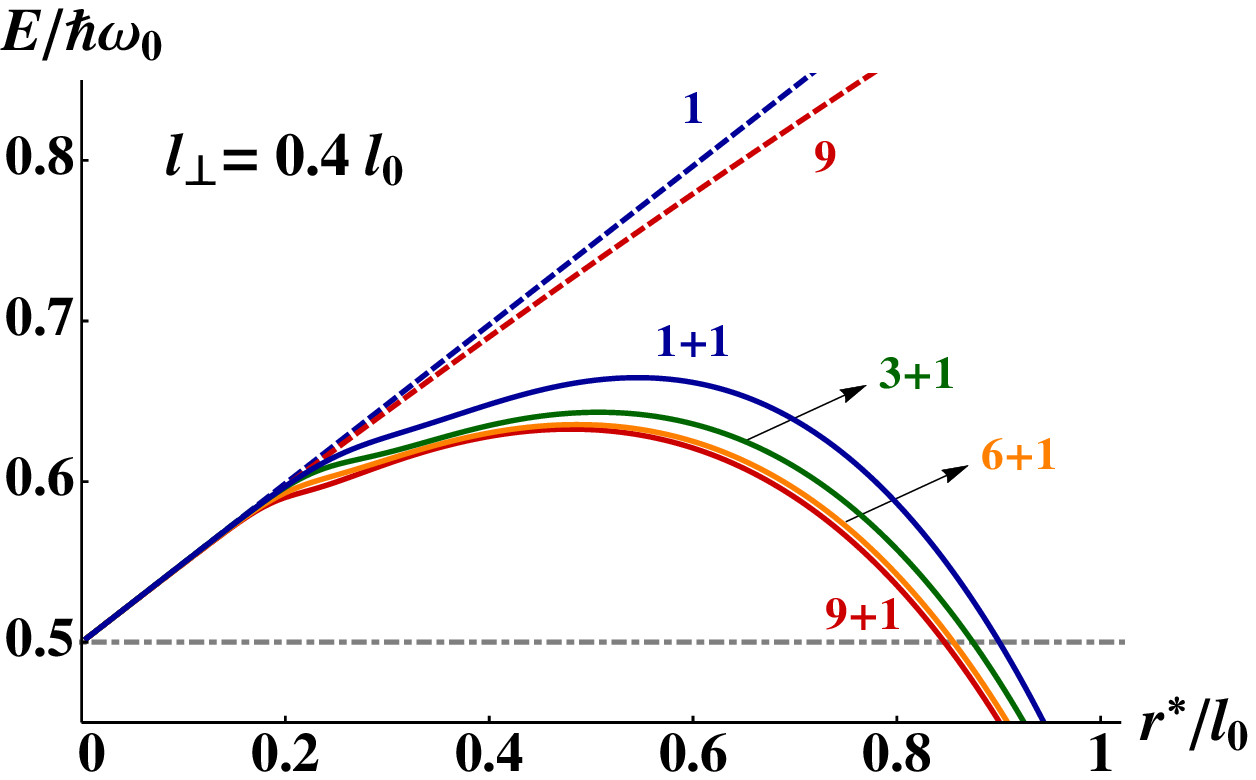}
  \end{center}
  \begin{center}
    \includegraphics[width=.31\textwidth]
    {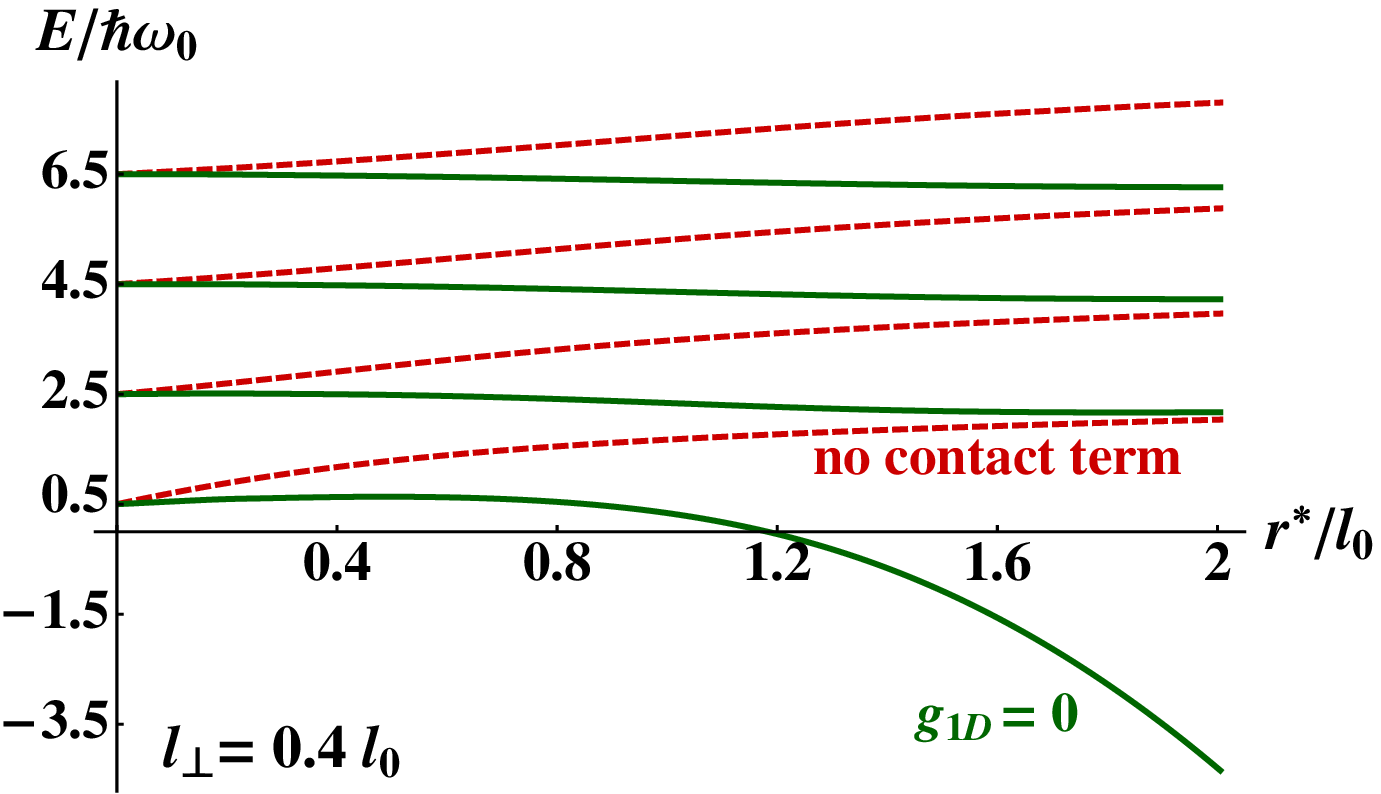}
  \end{center}
  \caption{ \label{fig:contact_nocontact} (Color online).
    Top: Ground--state energy $E_\mathrm{2B}(r^*)$ of the 
    Hamiltonian $H_\mathrm{2B}$,
    for $g_{1D}=0$ and $l_\perp/l_0=0.4$, 
    including $1$ (blue), $3$ (green), $6$ (orange), and  $9$ (red) harmonic
    oscillator states,
    without (dashed lines) and with (solid lines) the `bare' 
    bound state $\ket{\phi_{-1}}$ in the projection basis.
    Bottom: the four lowest eigenvalues of $H_\mathrm{2B}$ 
    as a function of $r^*$, for $g_{1D}=0$ (green) 
    and choosing 
    $g_\mathrm{1D}={2\hbar^2r^*}/{(3m l_\perp^2)}$ 
    to cancel the contact term (dashed red),
    calculated including six harmonic oscillator states and
    the `bare' bound state $\ket{\phi_{-1}}$ in the basis.
  }
\end{figure}

The applicability of the quasi--1D effective potential (Eq.~(\ref{eq:V1D})) to our harmonically confined system requires $l_\perp/l_0$ to be small. 
The energy $E_\mathrm{2B}(r^*)$ is plotted in Fig.~\ref{fig:twostate}(bottom) for $l_\perp/l_0=0.4$. Including more harmonic oscillator states in the basis allows for the calculation of higher--energy states (Fig.~\ref{fig:contact_nocontact} (bottom)). However, it does not affect the qualitative behavior of the ground--state energy $E_\mathrm{2B}(r^*)$ as long as $\ket{\phi_{-1}}$ is also included (Fig.~\ref{fig:contact_nocontact}(top)).

The non--monotonic behaviour of $E_\mathrm{2B}(r^*)$ 
is a signature of the DIR. The ground state energy goes below $\hbar\omega_0/2$ for $r^*>r^*_\mathrm{crit}$, where $r^*_\mathrm{crit}/l_0=0.90$ for $l_\perp/l_0=0.4$. 
In the many--body treatment described below, we are interested in situations where the dimer population is very small.
Similarly to Feshbach resonance physics \cite{chin_RMP2010}, the existence of the closed channel has a strong impact even though it is only marginally populated. Moreover, the dimer population being nearly vanishing  will help us simplify the problem to an effective open--channel model.
This assumption is satisfied here, as 
the overlap $|\braket{\phi_{-1}}{\Psi_{2B}}|^2$ remains smaller than $0.10$ for $r^*\lesssim r^*_\mathrm{crit}$. This overlap only becomes substantial if $\ket{\phi_0}$ and $\ket{\phi_{-1}}$ have comparable energies, i.e.~for
$r^*/l_0 \gtrsim 2.13$ (Fig.~\ref{fig:twostate}(top)). 
The bound state population near $r^*_\mathrm{crit}$ increases as $l_\perp/l_0$ decreases, but it remains $<0.15$ for $l_\perp/l_0\geq 0.2$.

Figure~\ref{fig:contact_nocontact}(bottom) shows the $r^*$--dependence of the lowest eigenvalues of $H_\mathrm{2B}$ 
in two different situations:
\textit{(i)} the $s$--wave interaction term $g_\mathrm{1D}=0$ and 
\textit{(ii)} $g_\mathrm{1D}\neq 0$ cancels the contact term in Eq.~(\ref{eq:V1D}) completely \cite{deuretzbacher_PRA2010}. 
The $r^*$--dependence of the energy levels in these two situations is completely different. This will allow for an observation of the DIR using spectroscopic techniques \cite{heckerDenschlag_JPB2002}.

\section{Many--body physics} 
We now consider $N$ dipolar particles in a deep quasi--1D optical lattice with unity filling factor. We describe this system using a Bose--Hubbard model \cite{bloch_NatPhys2012,fisher_PRB1989} extended to include nearest--neighbour interactions. We focus on the regime $r^*\lesssim r^*_\mathrm{crit}$, so that the DIR affects the two--body properties even though the number of dimers present in the system is extremely small. In order to properly account for the resonance, we start from the two--state description introduced above for the two--body problem (Eq.~(\ref{eq:twostate})). Each of the two states $\ket{\phi_0}$ and $\ket{\phi_{-1}}$ yields a band and, hence, we introduce an atom--dimer EBHM whose Hamiltonian reads:
\begin{multline} \label{eq:atdimEBHM}
H_\mathrm{AD}=
\sum_i 
\left[
  \varepsilon_a n_i 
  +\frac{U}{2} n_i (n_i-1) 
  -J_a( a_i^\dagger a_{i+1} + \mathrm{hc})
  \right. \\ \left.
  +V n_i n_{i+1} 
  +\varepsilon_d  m_i 
  -J_d( b_i^\dagger b_{i+1}+ \mathrm{hc})
  +\Omega (b^\dagger_i a_i a_i+ \mathrm{hc}) 
\right]
\ .
\end{multline}
In Eq.~(\ref{eq:atdimEBHM}), $a_i^\dagger$ and $b_i^\dagger$ are the creation operators in the site $i$ for atoms and dimers, respectively, and $n_i=a_i^\dagger a_i$ and $m_i=b_i^\dagger b_i$ are the corresponding number operators. Atoms and dimers are created in the ground state of the well $i$.
The atomic tunneling coefficient $J_a$ is taken from \cite{gerbier_PRA2005}.
The atomic on--site and nearest--neighbour interaction parameters $U$ and $V$ are defined in terms of $V_{1D}$ and the Wannier wavefunctions $w_i(x)$ and $w_{i+1}(x)$ localized on the sites $i$ and $i+1$ by \cite{trefzger_JPB2011}:
\begin{subequations} \label{eq:UVwannier}
  \begin{align}
  U &=\iint dx_1 dx_2\:
  w_i^2(x_1) w_{i}^2(x_2) \: V_\mathrm{1D}(x_1-x_2)
  \ ,
  \\
  V &=\iint dx_1 dx_2\:
  w_i^2(x_1) w_{i+1}^2(x_2) \: V_\mathrm{1D}(x_1-x_2)
  \ .
  \end{align}
\end{subequations}
We use the Gaussian approximation to the Wannier functions $w_i(x)$.
The on--site energy for atoms and dimers, $\varepsilon_a$ and $\varepsilon_d$, the atomic on--site interaction energy $U$, and the atom--dimer conversion rate $\Omega$, can then all be expressed in terms of the matrix elements appearing in Eq.~(\ref{eq:twostate}), namely:
\begin{subequations}
  \begin{align}
    \varepsilon_a &=\frac{1}{2}\hbar\omega_0 
    \ ,\\
    \varepsilon_d &=\varepsilon_a+
    \bra{\phi_{-1}}H_\mathrm{2B}\ket{\phi_{-1}}
    \ ,\\
    U &=\bra{\phi_0} H_\mathrm{2B} \ket{\phi_0}-\varepsilon_a
    \ ,\\
    \Omega &=\frac{1}{\sqrt{2}}\bra{\phi_{-1}}H_\mathrm{2B}\ket{\phi_0}
    \ .
  \end{align}
\end{subequations}
The nearly--vanishing dimer population allows for a crude description of the dimer dynamics, therefore we neglect atom--dimer and dimer--dimer interaction, and we take $J_d=J_a/10$ \footnote{This choice for $J_d$ reflects the assumption that the polarizability of a molecule is twice that of an atom \cite{gerbier_PRA2005}. In the considered regime, its exact value does not affect our numerical results.}. 

We focus on the tight--binding regime and we introduce the harmonic oscillator length $l_0$ characterizing the bottom of each lattice well. Like for the two--body problem, we consider a fixed and small value of $l_\perp/l_0$.
The ground state of the system then depends on two adimensional parameters: $r^*/l_0$ and $V/U$. The choice of the parameter $r^*/l_0$  allows for a direct comparison with the two--body physics illustrated in Figs.~\ref{fig:twostate} and \ref{fig:contact_nocontact}.  Assuming $g_{1D}=0$, Eqs. (\ref{eq:UVwannier}) show that
the ratio $V/U$ does not depend on $r^*$. It decays
with the lattice depth $s=V_\mathrm{lat}/E_R$, where $V_\mathrm{lat}$ is the intensity of the optical lattice and $E_R$ is the recoil energy. The harmonic approximation requires $s$ to be large enough
and thus imposes an upper bound on $V/U$.

\begin{figure}
  \begin{center}
    \includegraphics[width=.3\textwidth]
    {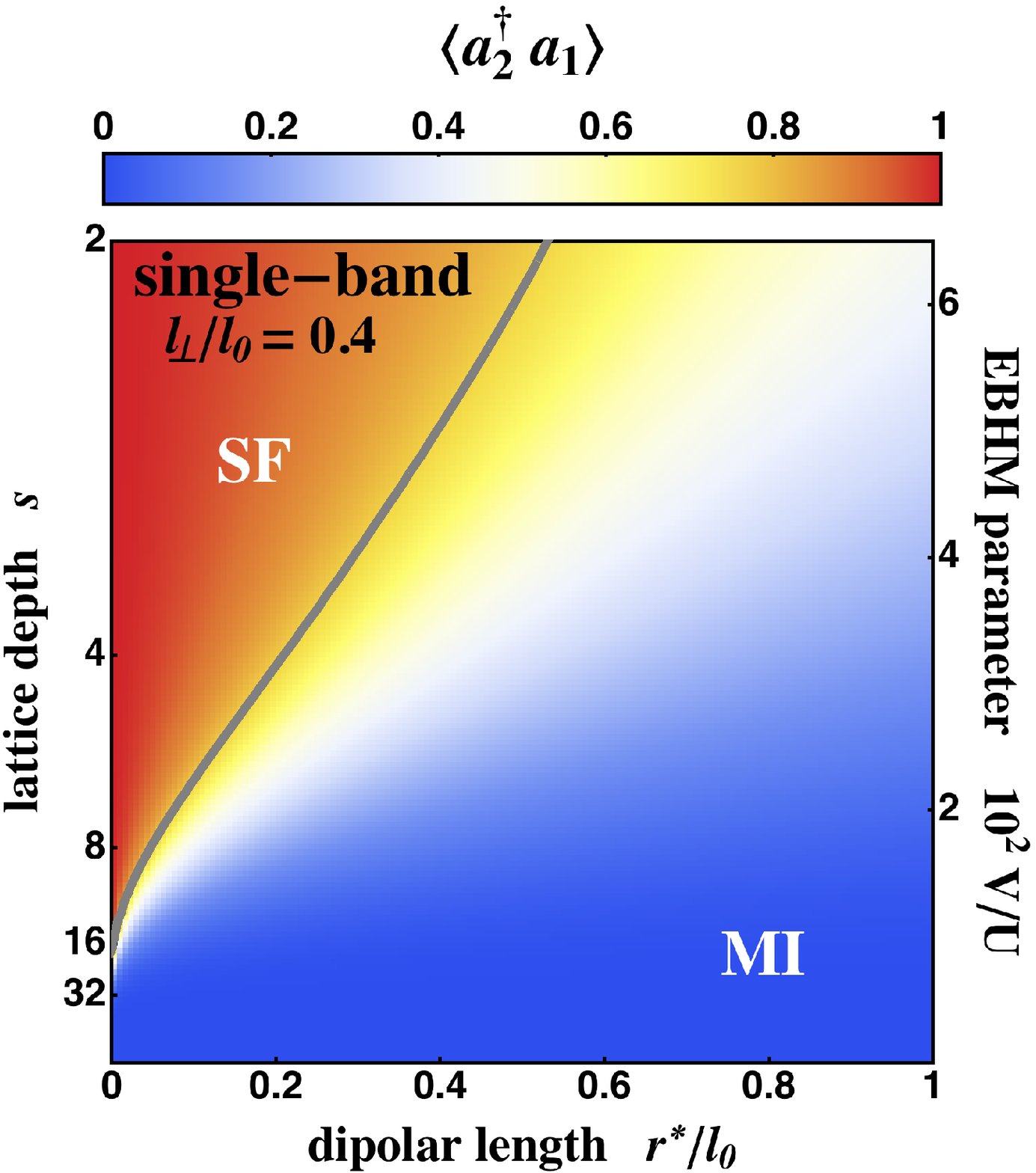}
  \end{center}
  \begin{center}
    \includegraphics[width=.3\textwidth]
    {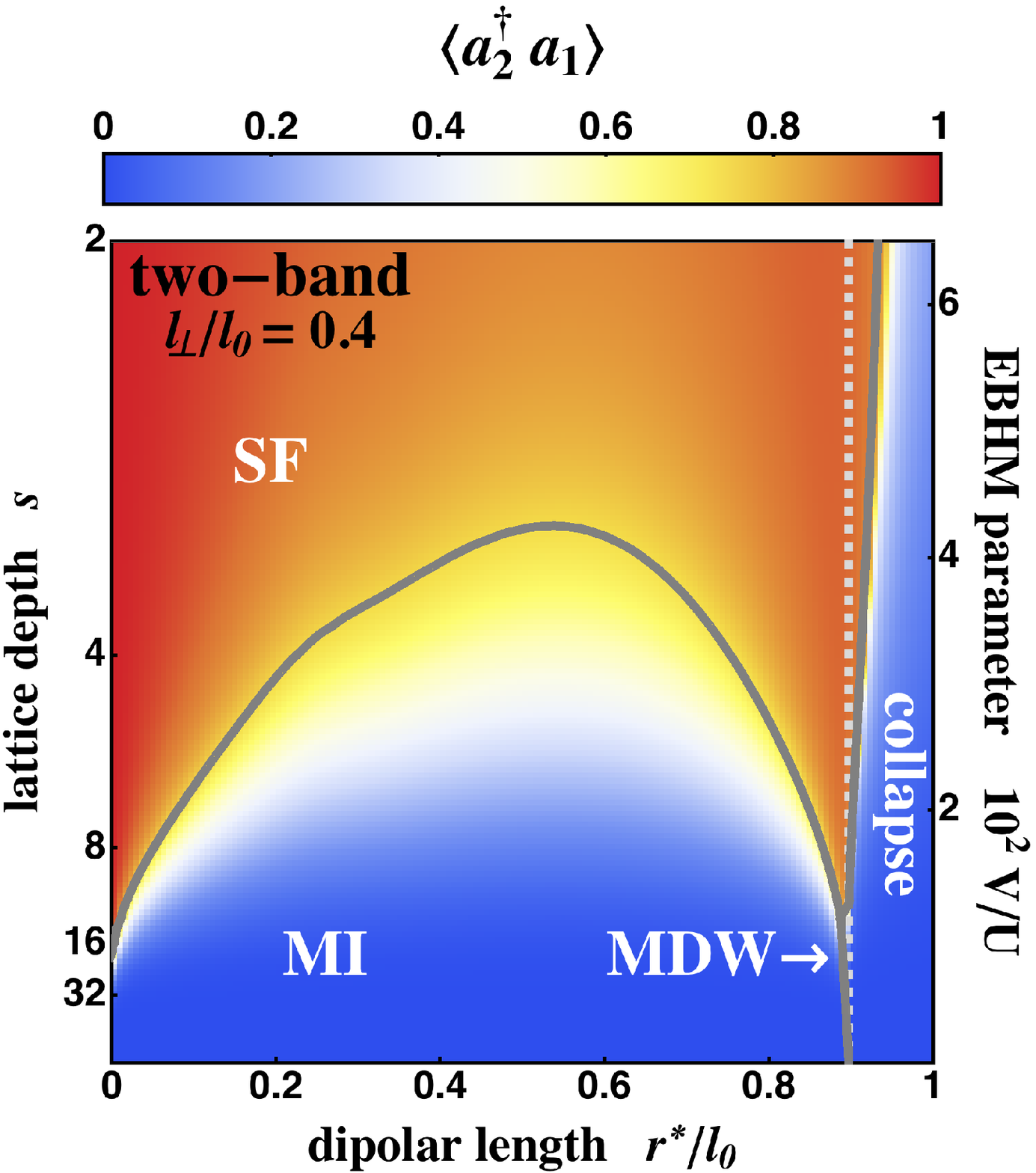}
  \end{center}
  \caption{ \label{fig:phasediags} (Color online). 
  Many--body phase diagrams obtained using the single--band ($\Omega=0$, top) 
  and atom--dimer ($\Omega\neq 0$, bottom) EBHMs, performing exact diagonalization
  on a six--atom, six--well system with $l_\perp/l_0=0.4$.
  The effective on--site interaction $U_\mathrm{eff}<0$ on the right of the 
  vertical dashed line.
  }
\end{figure}
For given values of $r^*/l_0$ and $V/U$, we numerically calculate the ground state of the $H_\mathrm{AD}$ by exact diagonalization of a $6$--atom, $6$--well system. 
Figure~\ref{fig:phasediags} shows the phase diagram of the system for $l_\perp/l_0=0.4$. The observable is the single--particle off--diagonal density matrix element 
$\rho_1=\langle a_{2}^\dagger a_1 \rangle$, 
and it distinguishes the superfluid phase ($\rho_1 \neq 0$) 
from the insulating phases ($\rho_1=0$). The different insulating phases can subsequently be told apart by examining the ground--state wavefunction. Figure~\ref{fig:phasediags} compares the physically accessible phase diagram obtained using the single--band EBHM \cite{pai_PRB2005}
(taking $\Omega=0$ in Eq.~(\ref{eq:atdimEBHM})) with the atom--dimer phase diagram ($\Omega\neq 0$).
In the considered range of parameters,
the previously investigated single--band phase diagram 
exhibits two phases: superfluid (SF) and Mott--insulator (MI). The atom--dimer phase diagram presents three qualitative differences. First, the MI phase region stops at $r^*=r^*_\mathrm{crit}$. Second, the phase diagram includes a narrow Mass--Density--Wave domain which occurs for very small values of 
$V/U$ \footnote{Up to now, the MDW phase had been predicted to occur in an extended domain corresponding to large values of $V/U$ \cite{pai_PRB2005}.}.
Third, there appears a `collapse' phase where all atoms sit in the same well
\footnote{This collapse phase is related to the one predicted in the 2D case \cite{goral_PRL2002} using a mean--field approach to calculate the Bose--Hubbard parameters.}
\footnote{Our phase diagram shows no phase with a period of three sites or more. We have checked that longer--period insulating phases are not energetically favored.
This is in agreement with the DMRG calculations including next--nearest--neighbor interactions reported in \cite{dallaTorre_PRL2006}.
However, such phases have been predicted to occur for filling factors $\neq 1$ (see e.g. \cite{burnell_PRB2009}).}.
In our small--sized system, the MI--MDW and MDW--collapse transitions appear sharp, in accordance with their expected first--order character. Instead, the transitions between the SF phase and each insulating phase are smooth, which is compatible with the  Berezinskii--Kosterlitz--Thouless behavior predicted in 1D \cite{pai_PRB2005}. 
\begin{figure}
  \begin{center}
    \includegraphics[width=.32\textwidth]
    {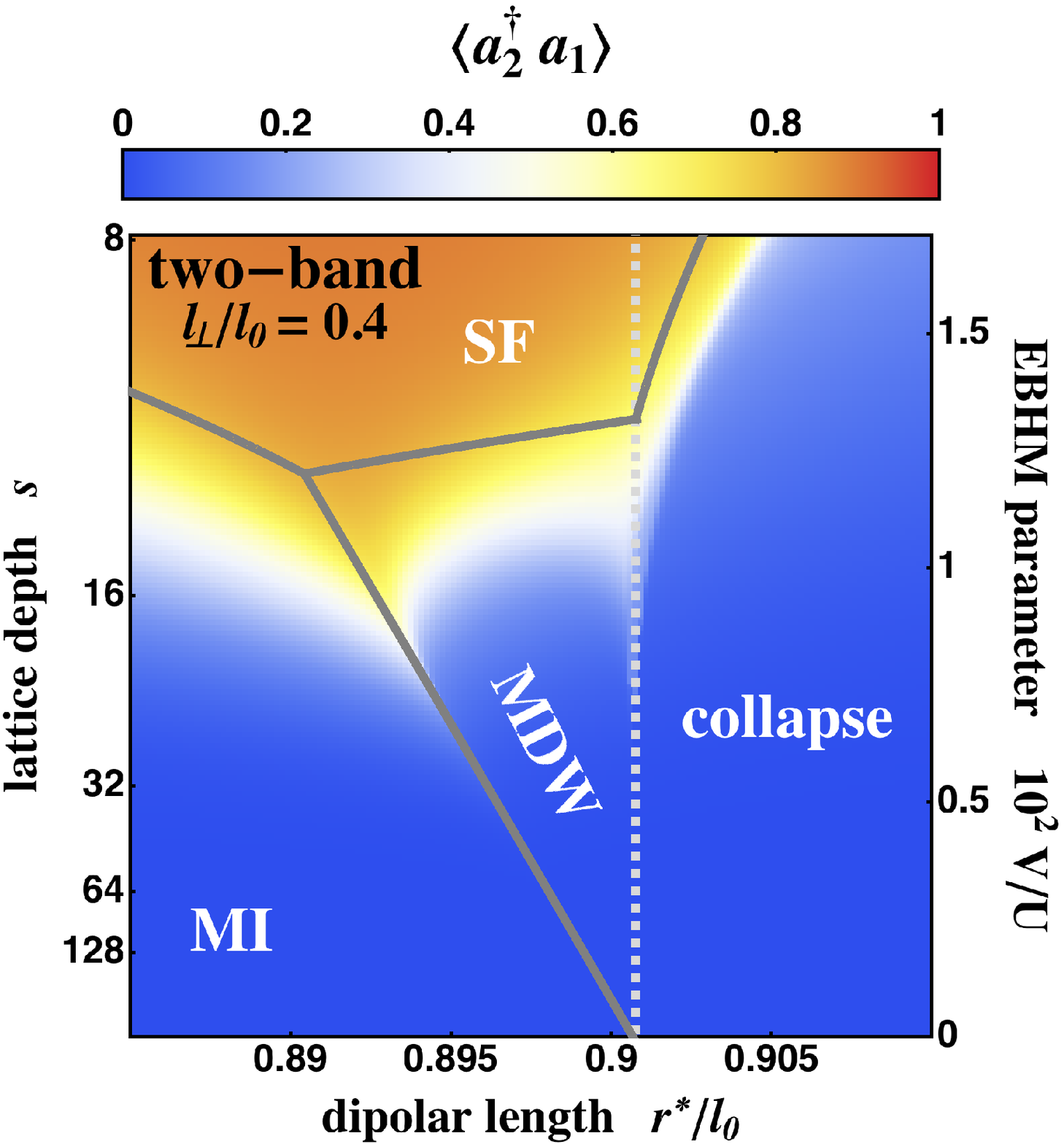}
  \end{center}
  \vspace{8pt}
  \begin{center}
    \includegraphics[width=.32\textwidth]
    {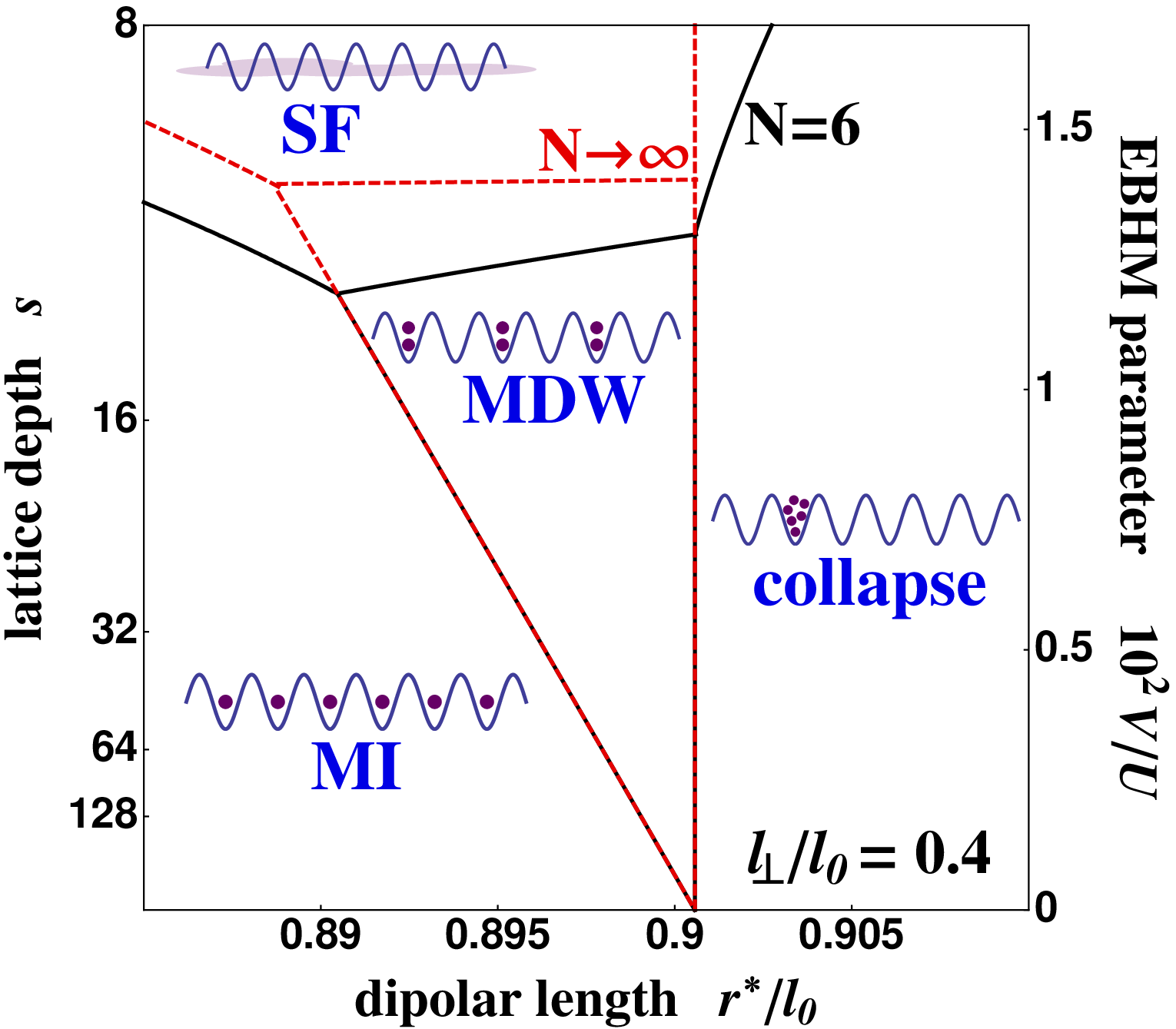}    
  \end{center}
  \caption{ \label{fig:phasediag_zoom_cartoon} (Color online). 
    Top: Zoom--in on the part of the atom--dimer phase diagram 
    (Fig. \ref{fig:phasediags} (bottom), $l_\perp/l_0=0.4$) showing the 
    transitions between the SF, MI, MDW, and collapse phases.
    Bottom: quasi--analytical phase boundaries calculated for $N=6$ 
    (solid black) and $N\rightarrow\infty$ (dashed red). 
  }
\end{figure}

\begin{figure}
  \begin{center}
    \includegraphics[width=.305\textwidth]
    {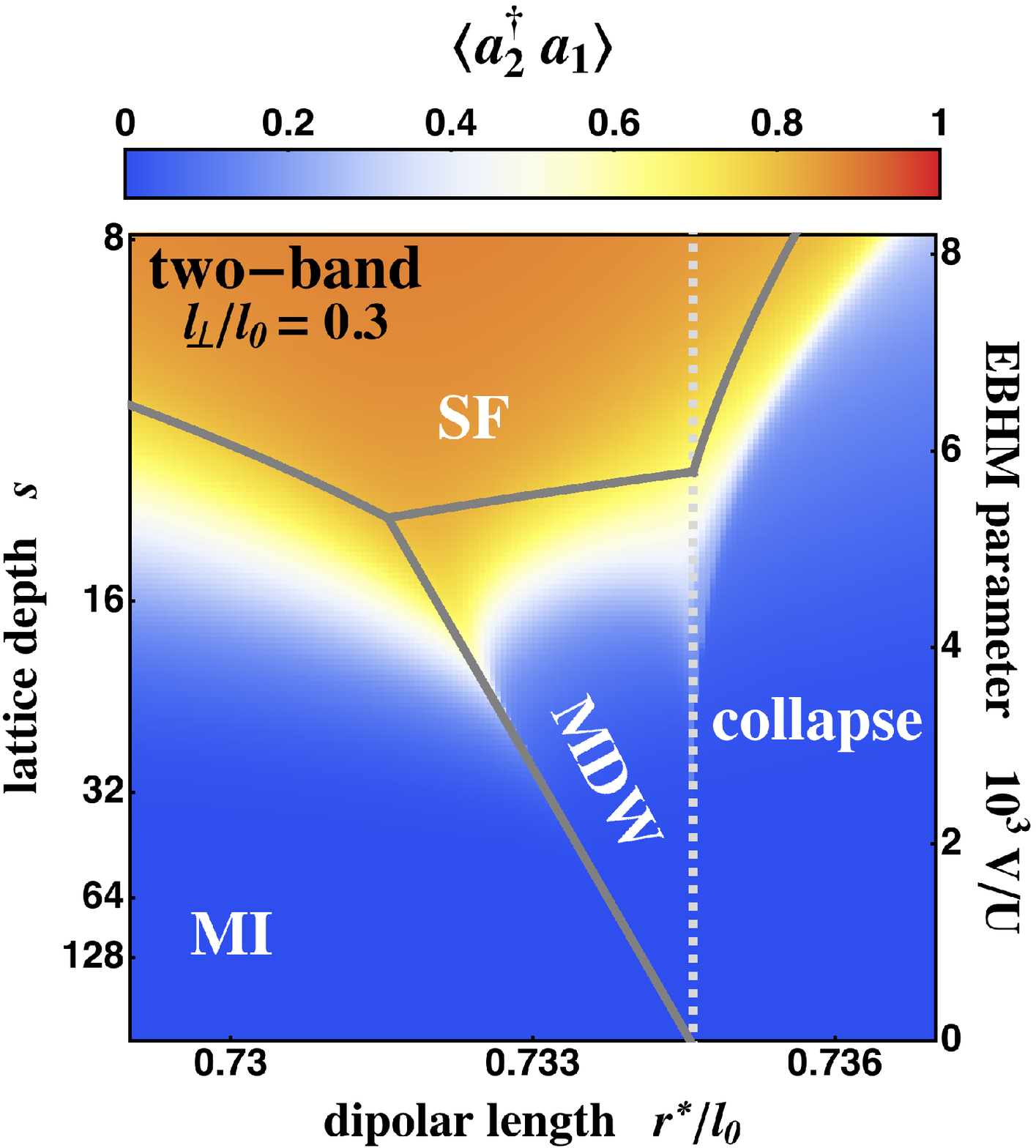}
  \end{center}
  \begin{center}
    \includegraphics[width=.305\textwidth]
    {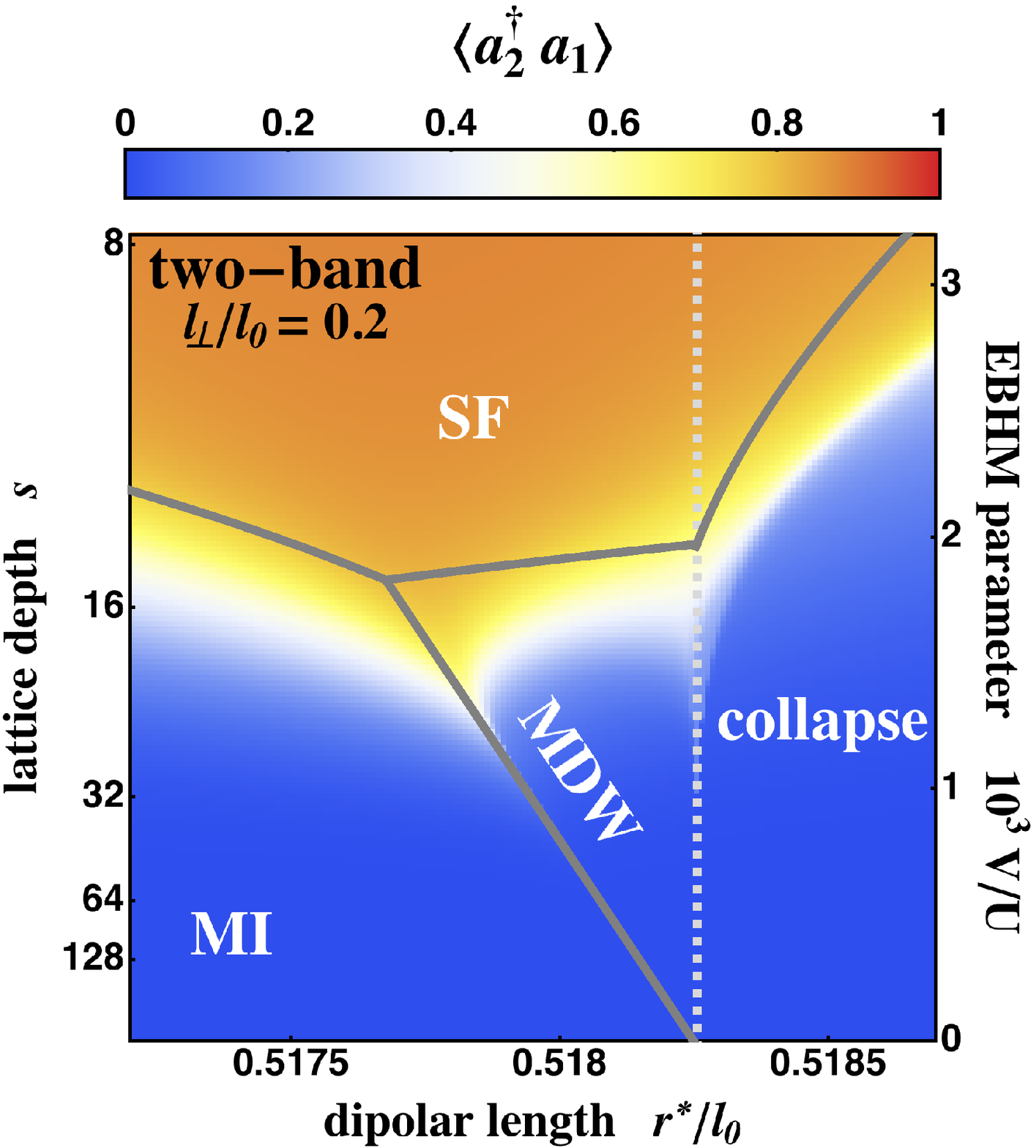}
  \end{center}
  \caption{ \label{fig:phasediag_zoom_0302}
    Zoom--in on the atom--dimer phase diagrams for $l_\perp/l_0=0.3$ (top) 
    and $l_\perp/l_0=0.2$ (bottom).
  }
\end{figure}

Figure \ref{fig:phasediag_zoom_0302} shows a zoom--in on the 
atom--dimer phase diagrams for $l_\perp/l_0=0.3$ and $0.2$. The comparison between Figs. \ref{fig:phasediag_zoom_cartoon} and \ref{fig:phasediag_zoom_0302} shows that decreasing the value of $l_\perp/l_0$ has a two--fold effect on the phase diagram:
\textit{(i)} the collapse phase, which starts at
$r^*=r^*_\mathrm{crit}$, appears for smaller values of $r^*/l_0$, and
\textit{(ii)} the extent of the MDW phase domain is reduced. This second result suggests that the experimental observation of  MDW phases in quasi--1D bosonic systems will be difficult.

The phase diagram can be interpreted using an effective single--band EBHM, where the on--site interaction reproduces the two--body ground--state energy:
\begin{multline} \label{eq:EBHMeff}
H_\mathrm{eff}\!\!=\!\!\!
\sum_i 
\left[
  -J_a( a_i^\dagger a_{i+1} + \mathrm{hc})
  +\varepsilon_a n_i 
  \right. \\ \left.
  +\frac{1}{2}U_\mathrm{eff} n_i (n_i-1) 
  +V n_i n_{i+1}
\right]
\ ,
\end{multline}
with $U_\mathrm{eff}(r^*)=E_\mathrm{2B}(r^*)-\varepsilon_a$. In the parameter range explored on Figs.~\ref{fig:phasediags} and \ref{fig:phasediag_zoom_cartoon}, the phase diagram obtained using $H_\mathrm{eff}$ is very similar to the atom--dimer phase diagram. This is due to the atom--dimer detuning $\Delta=\varepsilon_d-U-2\varepsilon_a$ being much larger than $\Omega$, $J_a$ and $V$ \footnote{The two approaches are expected to yield different results for small $\Delta$. Then, the dimer population is non--negligible, and the physics is described by the atom--dimer model of Eq.~(\ref{eq:atdimEBHM}) where the parameters modeling the dimer dynamics should be properly chosen.}. The effective model $H_\mathrm{eff}$ allows for a comparison between our phase diagram and those calculated in terms of the EBHM parameters $U/J$ and $V/J$. In particular, we find a Haldane--like phase near the upper left corner of our MDW domain, in agreement with the Haldane domain reported in \cite{rossini_NJP2012} 
\footnote{A systematic investigation of the Haldane phase domain will be carried out on larger systems using DMRG.}.

We also use the effective single--band model to derive quasi--analytical approximations for the phase boundaries for any number $N$ of particles and sites. We calculate the energy deep within each phase in terms of $J$, $U_\mathrm{eff}$, $V$, and $N$. Equating these energies for two contiguous phases, we obtain the boundaries shown for $N=6$ on Figs.~\ref{fig:phasediags} and \ref{fig:phasediag_zoom_cartoon}, and for $N\rightarrow\infty$ on Fig.~\ref{fig:phasediag_zoom_cartoon}(right).  The boundaries found for $N=6$ and for $N\rightarrow\infty$ are very similar. 
We now focus on the boundary between the SF and the collapse phases, given by
$E_\mathrm{SF}-E_\mathrm{collapse} \approx N(V-2J)-N^2U_\mathrm{eff}/2=0$. 
The tunneling term scales with $N$, whereas the interaction scales with $N^2$. Hence, for small $N$, the superfluid phase survives in a region where $U_\mathrm{eff}<0$, but the collapse phase is energetically favored for large $N$. This instability 
corresponds to the implosion of a Bose--Einstein condensate with a negative scattering length when its size is increased \cite{donley_Nature2001}. 

\section{Outlook}

The phase diagram which we have obtained describes the ground state of the system. One possible way to explore it experimentally is to cool the system in a given geometry in the absence of dipolar interactions, and then adiabatically increase $r^*$. The phases we have predicted at $T=0$ may be experimentally identified using in--situ imaging techniques as well as the recent advances allowing for the detection of non--local order \cite{endres_Science2011}.
The narrow MDW domain which appears in the phase diagram for small $V/U$ is well within the validity range of our atom--dimer Hamiltonian.
By contrast, the MDW phase domain previously
predicted using the standard EBHM \cite{pai_PRB2005,rossini_NJP2012} occurs in an extended domain corresponding to large values of $V/U$.

The DIR could also have a strong effect on systems described by generalized EBHMs such as the one studied in \cite{sowinski_PRL2012}. It would be interesting to extend this work to 2D geometries, where the anisotropy of the dipolar interaction is expected to play a role. Our analysis would also be relevant for the understanding of the fermionic 1D EBHM with repulsive interactions, where the relevant phases are the Spin Density Wave (SDW), the Charge Density Wave (CDW), and the Bond Order Wave (BOW) \cite{ejima_PRL2007,nakamura_PRB2000,bhongale_PRL2012,barbiero_TBP2013}.

\emph{Acknowledgments.} We acknowledge stimulating discussions with M.~Baranov, G.~Ferrari, L.P.~Pitaevskii,
G.V.~Shlyapnikov, S.~Stringari, and W.~Zwerger.
L.B.~ acknowledges the hospitality of the BEC Center, Trento, during the initial stage of the project. 
This work has been supported by ERC through the QGBE grant and by Provincia Autonoma di Trento.

\end{document}